# Self-Replicating Mechanical Universal Turing Machine

**Ralph P. Lano**
Technische Hochschule Nürnberg - Georg Simon Ohm
Keßlerplatz 12, 90489 Nürnberg, Germany
ralph.lano@th-nuernberg.de

**Abstract**
This paper presents the implementation of a self-replicating finite-state machine (FSM) and a self-replicating Turing Machine (TM) using bio-inspired mechanisms. Building on previous work that introduced self-replicating structures capable of sorting, copying, and reading information, this study demonstrates the computational power of these mechanisms by explicitly constructing a functioning FSM and TM. This study demonstrates the universality of the system by emulating the UTM(5,5) of Neary and Woods.

.

## Introduction

In a previous study [1,2], we introduced a bio-inspired mechanism capable of sorting, copying, and reading; that is, it can perform basic information processing tasks. The existence of finite-state machines [16] and Turing machines [4,5] in biological and chemical contexts has been discussed in the literature [9-12]. In this study, we focus on this aspect of our mechanism, aiming to explicitly demonstrate its computational power.

In [3], we proposed a self-replicating Turing machine (TM). Although the design worked in principle, it was highly inefficient and impractical. Several details remained unclear, such as the glue mechanism, which was not clearly defined, and the issue of sorting blocks. In addition, the absence of a move-by-n component makes the design rigid and inflexible. These shortcomings were addressed in [1], where we clarified the missing details. In [2], we introduced an alternative mechanism inspired by protein folding in biology, which achieved the same results as in [1]. Both [1] and [2] focused on the building/self-replication mechanisms, whereas this study highlights the computational potential of these mechanisms.

In [1], we defined basic building blocks, including codons, tRNA, and the encoding of information in an RNA tape. In addition we, developed machines constructed from these basic building blocks, each of which performs specific functions. For example, the Copier machine was designed to copy the RNA tape, the Decoder machine was used to read information from the tape, and the Builder machine utilized this information to construct other structures and machines. Thus, when we refer to machines in this study, we mean the machines described in [1] and [2], built from these basic building blocks.

For brevity, we do not reintroduce the notation defined in [1-3].

## Finite-State Machine

A finite-state machine starts out in a given internal state, reads some symbol from some input, and based on that symbol, might change its internal state. There are a finite number of states and a finite number of symbols. A typical finite-state machine has its symbols arranged on some tape. However, the input could also come from some sensors. It also has a head for reading these symbols. The operation of such a machine is as follows: the machine starts out in a given internal state. In a first step, it reads a symbol from the tape at the current position of its head. Depending on a set of rules described in the state-transition table, the machine either switches its internal state to a new state or remains in the same state. Finally, the head is moved to the next position, and the process starts all over again.

The Decoder presented in [1] with minor modifications can function as a finite-state machine. To see this, let us consider an example of a finite-state machine that acts as a paritity detector. It detects whether the numbers of 1's in a given input string is even or odd. Figure 1 shows its state diagram. We start out in state A, that is, in the even state. We then start reading one symbol at a time. If the symbol read is 0, we remain in state A. If it is 1, then we switch to state B, which indicates an odd number of 1's. While in state B, if we read 0, then we remain in state B. Only when reading 1, we switch back to state A.

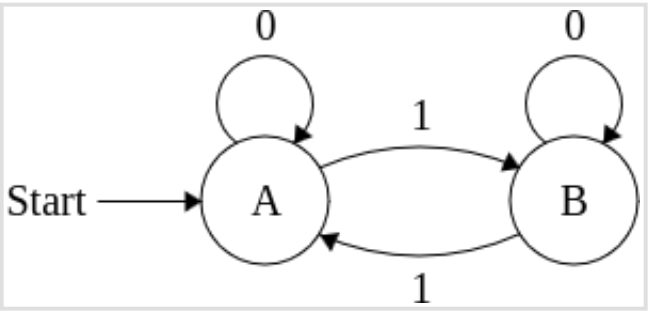

a

Figure 1: State diagram for parity detector

Usually, a finite-state machine is described via its state-transition table, as shown in Table 1. The input symbols are in the left-most column, and the current state the machine is in before reading is depicted in the top row. The entries in the center indicate the new state after reading a given symbol.

| | A | B |
|---|---|---|
| 0 | A | B |
| 1 | B | A |

Table 1: State-transition table for parity detector

Our goal is to use the Decoder presented in [1] as a state machine. For this purpose, it is more convenient to rewrite the state-transition table in the form of a lookup table (Table 2), that maps the current state and a given input symbol to the new state. Table 1 and Table 2 are equivalent.

| Rule # | Current state | Current input | New state |
|---|---|---|---|
| 1. | A | 0 | A |
| 2. | A | 1 | B |
| 3. | B | 0 | B |
| 4. | B | 1 | A |

Table 2: Lookup table for parity detector

Figure 2 shows one possibility of using 2-codons to encode the symbols '0' (Fig. 2(a)) and '1' (Fig. 2(b)), as well as a tape representing the input string '110'. We used blue color for symbols and dark gray color for the separators.

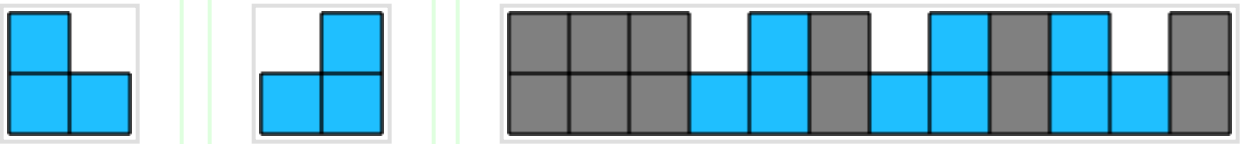

a b c

Figure 2: Symbols and tape

As for the state we use yellow color. Figure 3 shows how to use 2-codons to encode the states 'A' (Fig. 3(a)) and 'B' (Fig. 3(b)). Fig. 3(c) shows the tape for the above state machine which is in the initial state A.

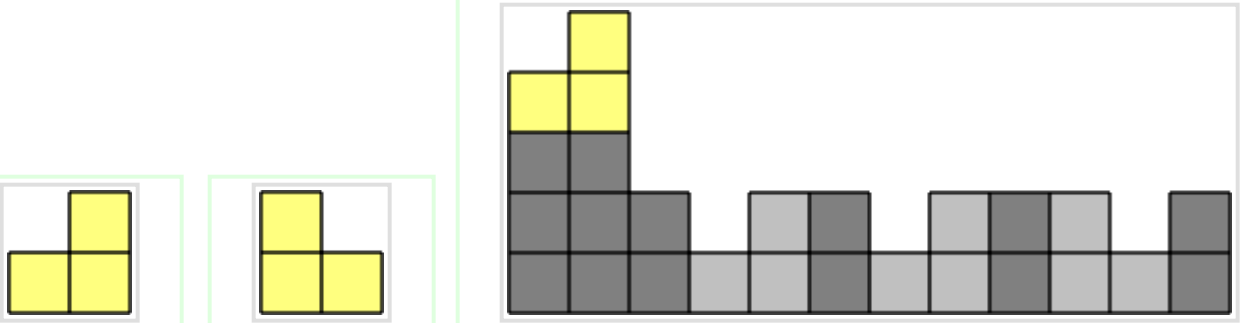

a b c

Figure 3: States and tape with initial state

We now use the state-transition table (Table 2) and translate it into tRNA.  It is basically a lookup table; for instance, the first line tells you that given state A and symbol 0, the new state will be state A.  The table suggests that there are four possible state transitions.

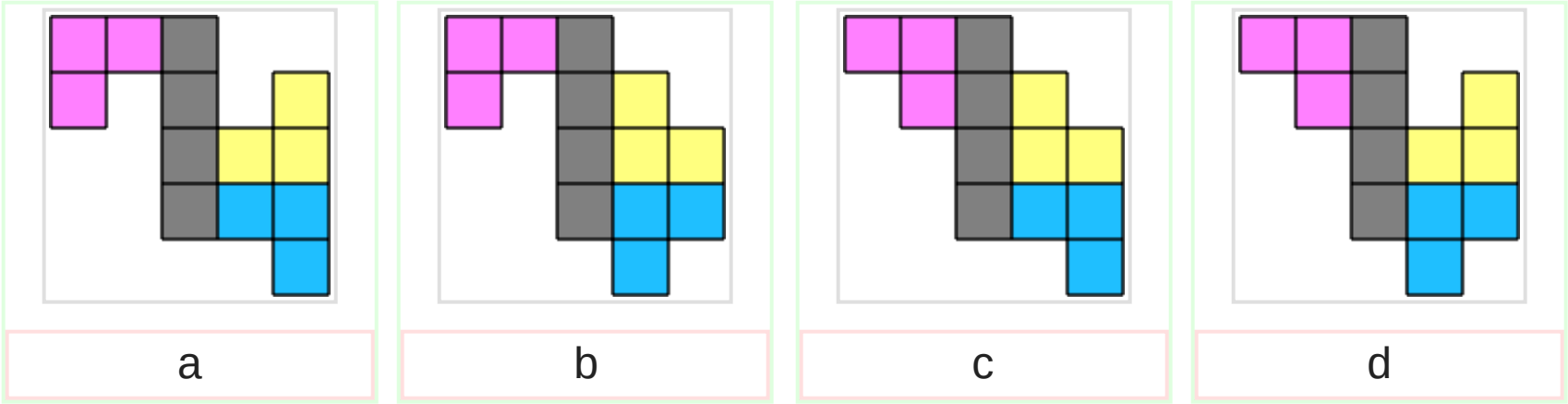

Figure 4: tRNA for state machine

The way to read this is the following.  The magenta part matches the current state.  The blue part matches the current symbol on the tape.  And the yellow codon is the new state.  With this understanding, we see that Fig. 4(a) corresponds to rule #1 of Table 2, Fig. 4(b) corresponds to rule #2, and so forth.  It is straight-forward to see, how any state transition table can be matched by corresponding tRNA.

Figure 5 shows the operation of the state machine.  We start with the tape and the initial state A, represented in yellow, and the next symbol to be read is symbol 1, as shown in Fig. 5(a).  The only tRNA that matches, is Fig. 4(b), that is rule #2, which is depicted in Fig. 5(b).  The new state is now state B.  Figs. 5(c) and 5(d) show the repetitive matching of tRNA and the operation of the state machine.  Fig. 5(d) shows the final result: the machine is in state A, the even state, because it has encountered two 1-symbols on the tape.

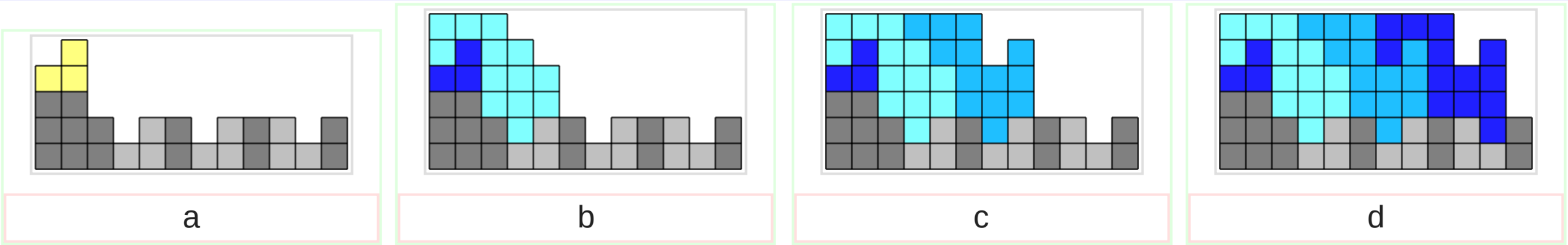

Figure 5: Operation of state machine

.

## Implementation

The Decoder machine presented and described in detail in [1], reads information from a tape of codons (RNA) by matching it with tRNA to create a stream of ordered blocks.  Interestingly, converting the Decoder into a state machine is straight-forward, as shown in Fig. 6.  We no longer needed to create a stream of ordered blocks; hence, we removed that part.  The conveyor needs to be positioned slightly lower, and instead of the move-by-two we use a move-by-three.

The input of the state machine is the tape from Fig. 5(a), shown in transparent, entering the machine from the left.  In addition, random tRNA from Fig. 4 enters the machine from above (Fig. 6(a)).  The tRNA is matched against the tape from behind (Fig. 6(b)).  Now, two things can occur:

a) There is a match: if it matches, it latches onto the tape, and the move-by-three Conveyor moves both the tRNA and the tape (Fig. 6(d)).
b) There is no match: if it does not match, the move-by-three Conveyor does nothing, the tape does not move, and the tRNA stays where it was.  In the next pass, this tRNA will be pushed out of the machine downwards.

Assuming a match, Fig. 6(f) shows the final result after one iteration.  The new state of the state machine is at the same location as the old state was, and the tape has been moved forward by one unit, ready to read the next symbol.

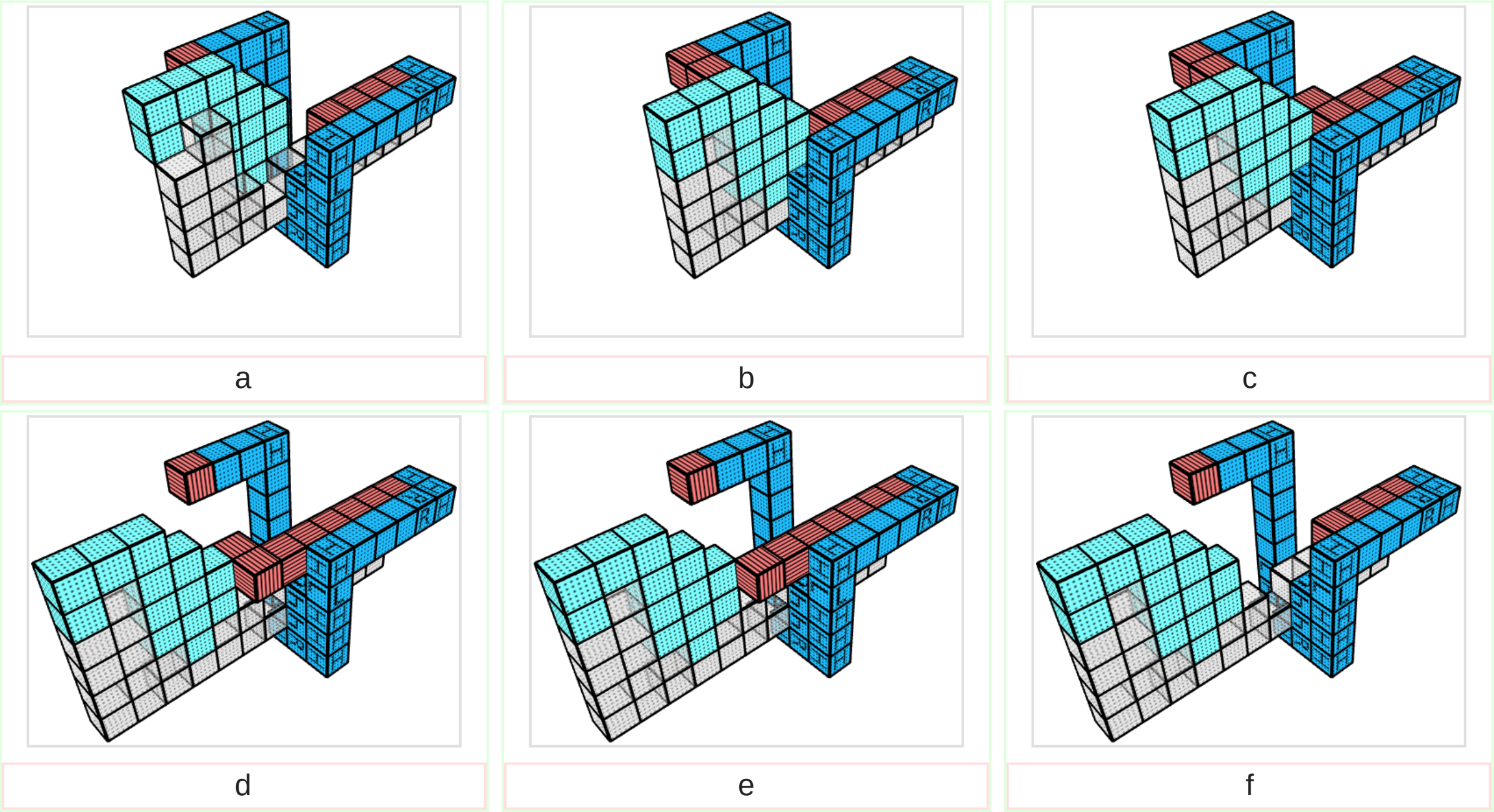

Figure 6: Decoder as state machine

The above state machine can be described using the machine description language (MDL) introduced in [2] as shown in Fig. 7.

```
M50b__b__H__b__b__b__b__L__H__L__h__H__R__H__H__h__h__H__L__
H__b__b__b__R__H__H__M32M32M32M41
```

Figure 7: MDL for state machine

Because we used the language from [2] to describe our state machine, it is clear that together with the mechanism from [2] this state machine can be built, and hence, the combination of the machines from [2], together with the state machine presented here, constitutes a self-replicating state machine.

## Generalization

The generalization to other state machines is straight-forward. Identify symbols and states, and select the correct size for codons. Table 3 depicts the number of symbols/states that can be represented with a given codon length (for details see [3]).

| n | C(n,n/2) |
|---|---|
| 1 | 1 |
| 2 | 2 |
| 3 | 3 |
| 4 | 6 |
| 5 | 10 |
| 6 | 20 |

Table 3: Codon length n depending on number of symbols/states

## Turing Machine

A Turing machine is basically a state machine that can not only read, but also write, and whose head can not only move in one direction, but in two. It is a state machine with two additional requirements: writing, and moving its head not only to the left but also to the right. The requirement that the tape be of infinite length is irrelevant for any physical Turing machine. To demonstrate how a Turing machine works, we start with a simple Turing machine, the binary incrementer shown in Fig. 8.

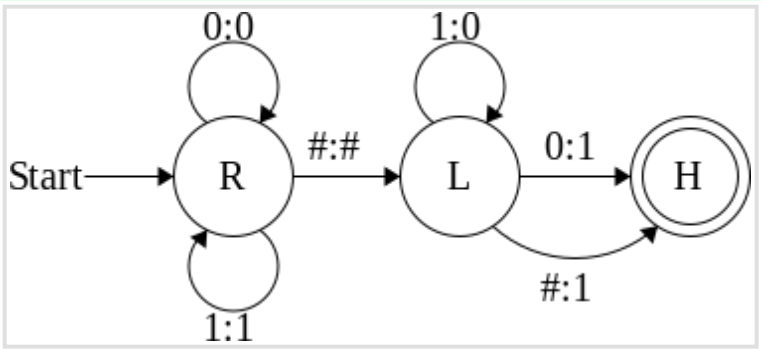

a

Figure 8: Binary incrementer

This Turing machine is uniquely described by its state transition table, as shown in Table 4. It first lists all the symbols and states, and then the rules for transitions between the states, depending on the current symbol read from the tape. For instance, rule #1 says that if we are in state R and read the symbol 0, then we are to stay in state R, write the symbol 0 to the tape at the current position, and then move the head to the right.

```
Symbols: 0, 1, #
States: R, L, H
Rules:
1.  (R, 0) → (R, 0)
2.  (R, 1) → (R, 1)
3.  (R, #) → (L, #)
4.  (L, 0) → (H, 1)
5.  (L, 1) → (L, 0)
6.  (L, #) → (H, 1)
```

Table 4: Transition table for binary incrementer

To convert this into our physical model, we start with the symbols, where Fig. 9(a) represents the '0', Fig. 9(b) represents the '1', and Fig. 9(c) represents the '#'.

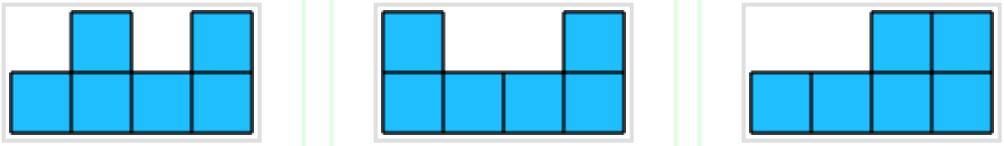

a b c

Figure 9: Symbols of TM

Then Fig. 10 shows the states, where Fig. 10(a) represents the 'R' state, Fig. 10(b) represents the 'L' state, and Fig. 10(c) represents the 'H' state.

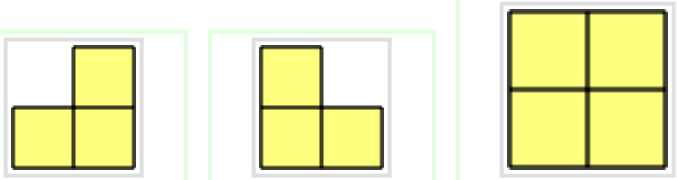

a b c

Figure 10: States of TM

Finally, the tape is shown in Fig. 11. The symbols are represented in light gray, whereas the states are represented in dark gray, with the current state depicted in yellow and the next symbol to be read in blue, to the right of the current state. Initially, the head of the machine is positioned above the blue symbol. Just considering the symbols, this tape corresponds to the input string '#01#'.

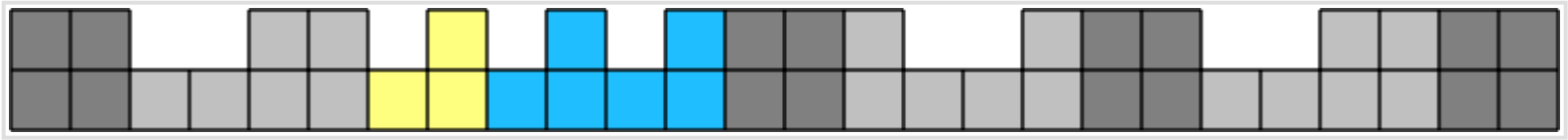

a

Figure 11: TM tape

In Fig. 12, we show the tRNA representing the transition table in Table 4. Here, Fig. 12(a) corresponds to rule #1, Fig. 12(b) corresponds to rule #2, etc.. As you can see, the tRNA consists of three parts: to the left and right in yellow, we have the state, and in the center, we have the symbol in blue. The lower part of the tRNA is used for reading and the upper part is used for writing.

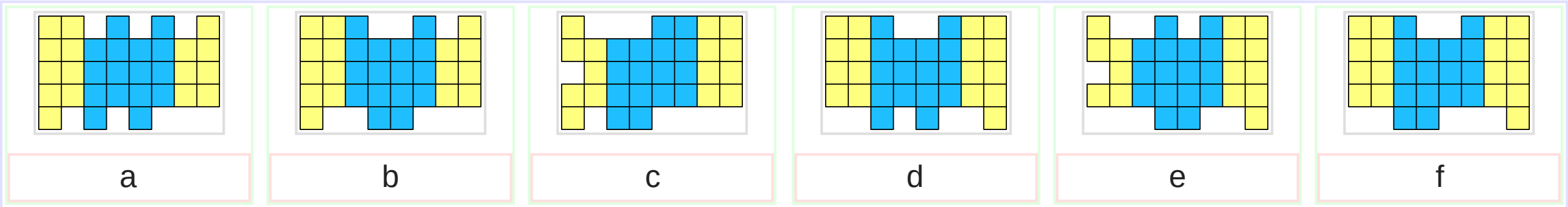

Figure 12: tRNA representing the transition rules

Figure 13 shows the operation of our Turing machine. Let us consider this in detail. Fig. 13(a) depicts the original tape, representing the input string '#01#', in the initial state 'R', and the head is centered over the second symbol (blue). The machine will then match randomly any of the six tRNA from Fig. 12 until it finds one that matches. Fig. 13(b) depicts a match, that is, the tRNA from Fig. 12(a). Once a match is found, that tRNA is then pushed down, as shown in Fig. 13(c), thereby replacing the old content of the tape with new content. This is the write operation. You may notice that the symbol remained unchanged, but the state has shifted to the right. Because R is a right moving state, we shift the head to the right by eight units, centering the head over the third symbol. Now we repeat the process of matching (i.e. reading), writing, and moving the head.

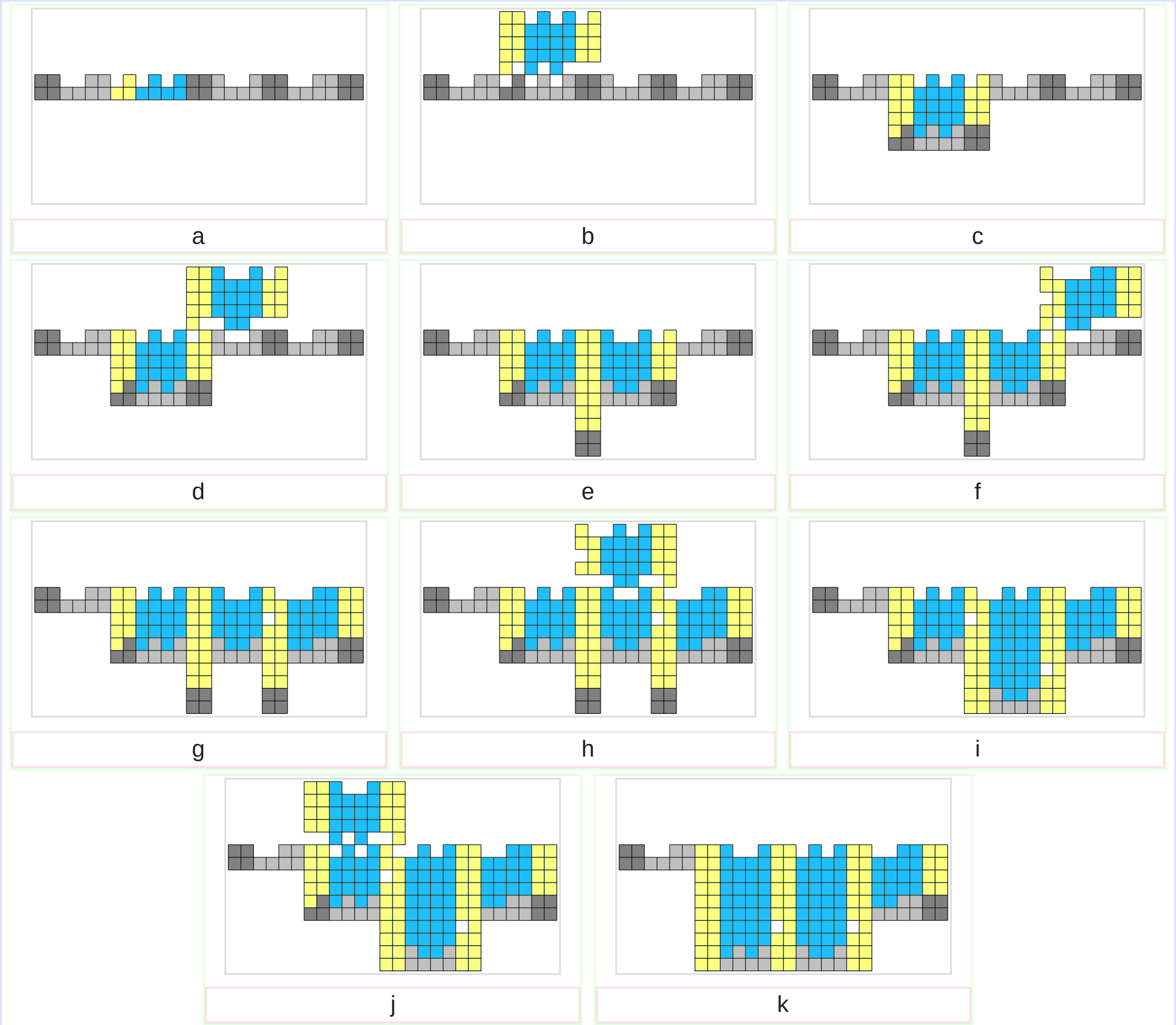

Figure 13: TM operation

Interesting is what happens in Fig. 13(f): here we shift from a right-moving state to a left-moving state. There are two things to observe: First, notice that the state information is no longer stored to the right of the symbol, but now to the left. This is because in the next move, our head moves to the left, and if we were to keep the state information to the right, we could no longer access the state information. Second, notice the small hole in the third row of the state information. All the left moving states have this hole. Later we will see how this is used by the machine to perform the actual left movement of the head.

Finally, let us observe Fig. 13(k): there is no state to the left or right of the symbol; rather, on both sides, we have the halting state. Because none of our tRNA will match this state-symbol pair, the machine will no longer move to the right or left; hence, it will halt.

.

## Implementation

To implement the above functionality in the form of a machine, we used the mechanism described in detail in [1]. Figure 14 shows the construction of the tape (Fig. 14(a)) and Turing machine (Fig. 14(b)).

In Fig. 14(a), the tape from Fig. 11 is recognized, enclosed by an encasing, shown in white. The encasing is needed because of the write operation, we can no longer glue the individual pieces of the tape together (that is the state and symbol codons). But rather we must be able to replace them. Since we still need to move the tape as a whole to the left or right, the encasing is required. Clearly, building an encasing for an infinite tape will be a little difficult.

This tape is enclosed by the Turing machine, as shown in Fig. 14(b). It consists of five active components. The first active component (1), shown in the middle, performs the matching; it simply attempts to match randomly entering tRNA against the tape. If there was no match, the active components (2) will remove the non-matched tRNA down, out of the way. In the case of a match, the move-left component (3) moves tRNA and tape to the left. If the tRNA has a hole, the move-right component (4) will attempt to move the tRNA and tape to the right. Finally, active components (5) and (6) push down, thus performing the write operation. All active components are connected through a support structure; however, for visualization purposes, the connections for components (2), (5), and (6) are not shown. However, for the machine to work, components (2), (5), and (6) must be connected to each other and to the other components.

The Turing machine shown in Fig. 14(b) works only for state-codons of length two and symbol-codons of length four. Clearly, this construction can be extended for other lengths; however, there is one constraint that this construction must require: states can be made up of odd and even codons, but for symbols, only even codons are allowed. This is because the conveyor can only move things by an even number.

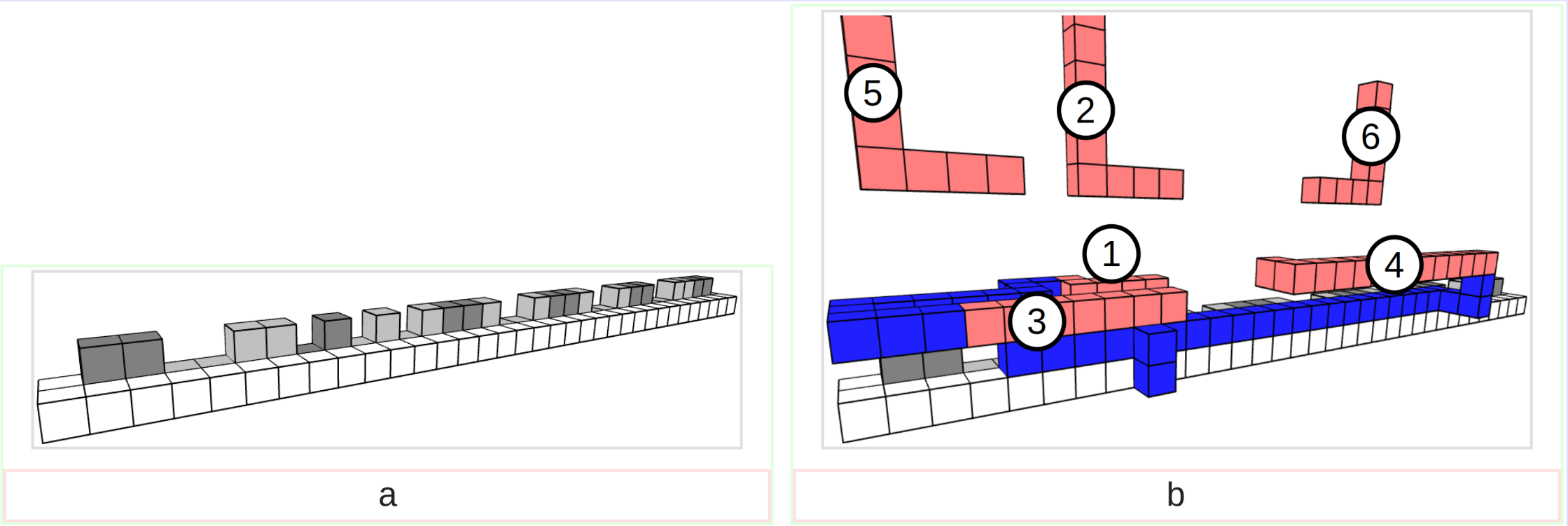

Figure 14: Tape and Turing machine

Figures 15-17 show our Turing machine in action. In Fig. 15 we depict the unsuccessful matching process. A random tRNA enters the machine from the right, and is under component (2), as shown in Fig. 15(a). As indicated above, component (1) attempts to match the tRNA against the tape (Fig. 15(b)), which is not successful. Finally, component (2) pushes downwards, thus removing the tRNA (Fig. 15(c)).

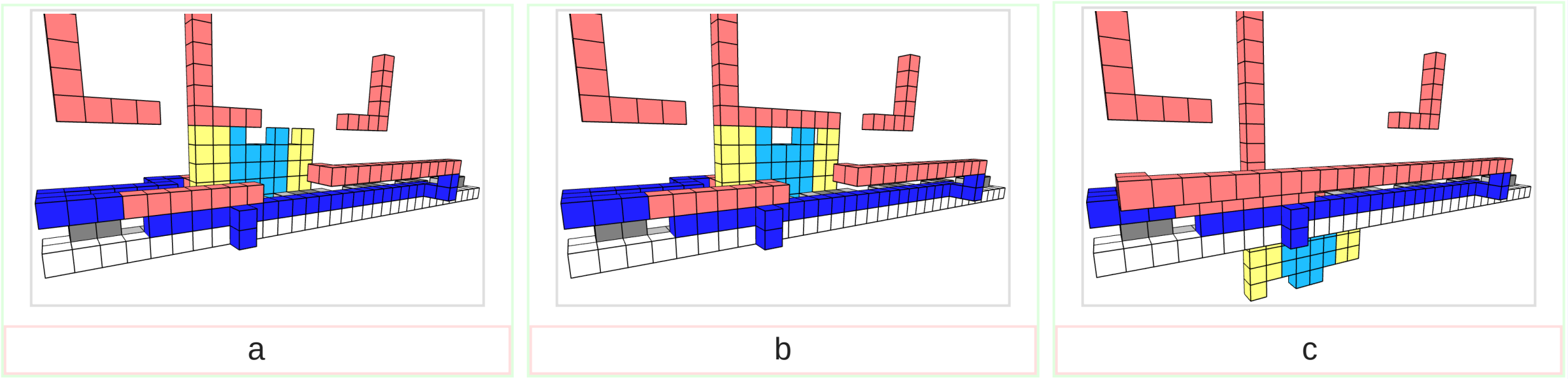

Figure 15: Turing machine: unsuccessful matching operation

Figures 16 shows the result of a successful match. We continue from Fig. 15(c). A new tRNA has entered the area below component (2), as shown in Fig. 16(a). Then, component (1) attempts to match the tRNA against the tape (Fig. 16(b)), which is successful. Next, component (3) grabs the tRNA (Fig. 16(c)) and moves it together with the tape, to the left (Fig. 16(d)). Figures 16(e) and 16(f) show how component (4) attempts to move tRNA and tape to the right, but because the shown tRNA does not have a hole to hold on to, is not successful. Figures 16(g) and 16(h) show the process of writing: here component (5) pushes down, thus replacing the old content of the tape with the new content, defined by the upper part of the tRNA. This corresponds to a normal operation, as shown in Fig. 13(a) through (c).

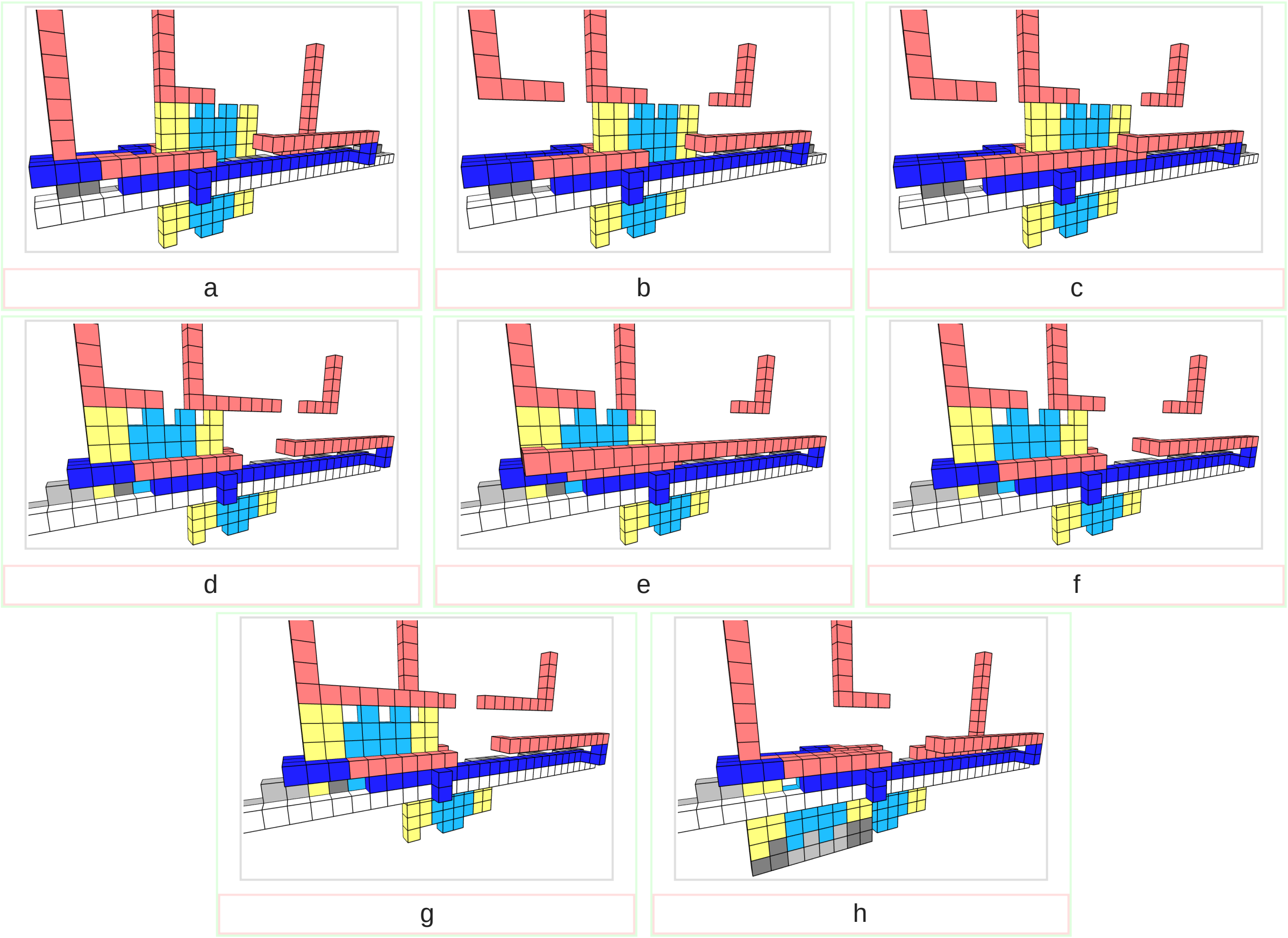

Figure 16: Turing machine: write operation, move to left

Figures 17 shows the result of a successful match, but resulting in a move to the right. A new tRNA has entered the area below component (2) and the match was successful, as shown in Fig. 17(a). Then, component (3) grabs the tRNA and moves it together with the tape, to the left (Fig. 17(b)). Next, Fig. 17(c) shows how component (4) can hold on to the tRNA by using the hole in the tRNA. Fig. 17(d) shows how component (4) then moves tRNA and tape to the right. Figures 17(e) and 17(f) show the process of writing: here component (6) pushes down, thus replacing the old content of the tape with the new content, defined by the upper part of the tRNA. Moving the tape to the left corresponds to moving the head to the right, corresponding to the process shown in Fig. 13(f) and (g).

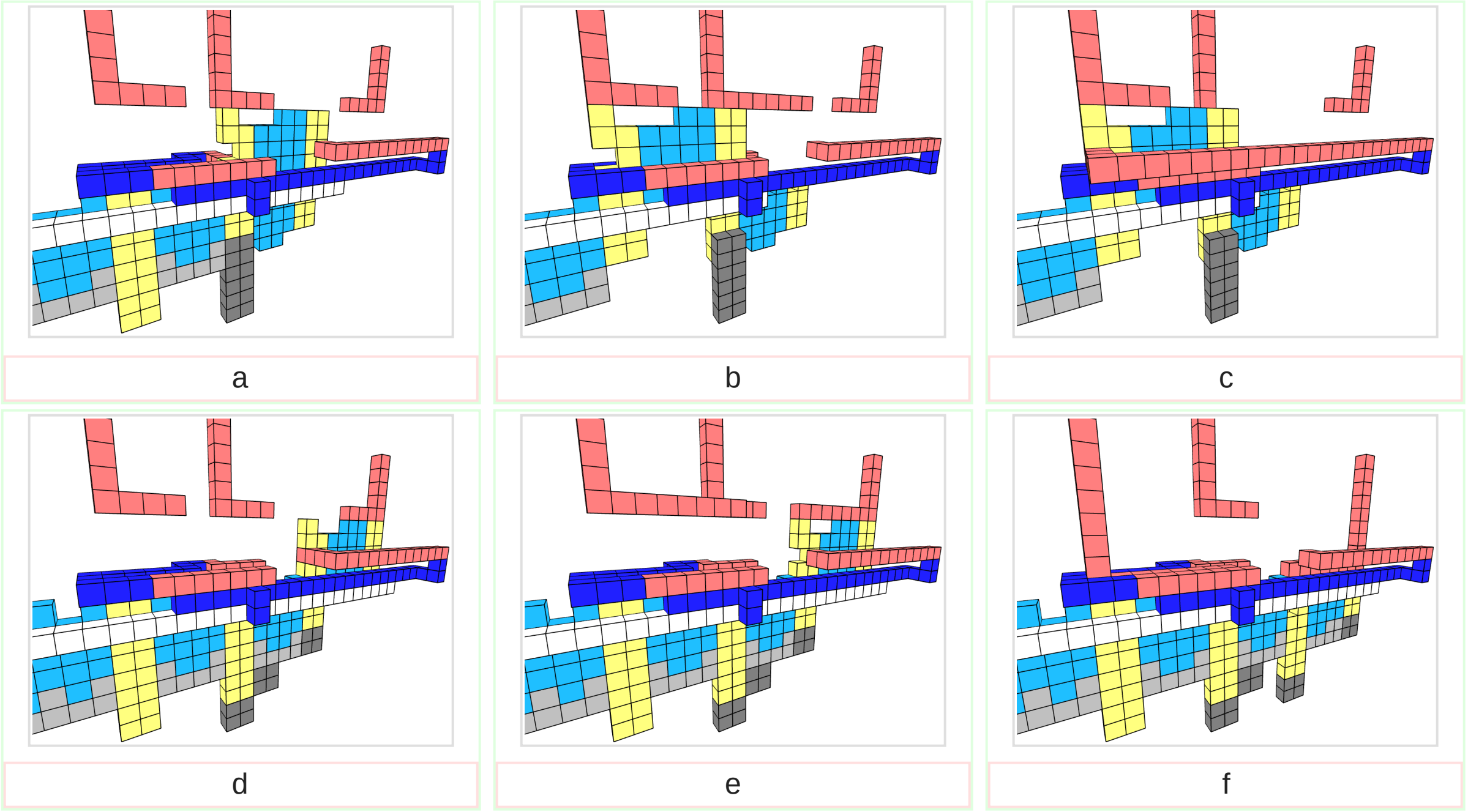

Figure 17: Turing machine: write operation, move to right

After these operations, the Turing machine will finally be in a state corresponding to Fig. 13(k), where all states are in the 'H' state, that is, the halt state. Because none of the tRNA will match, our machine will halt.

Appendix C provides a detailed description of the shown Turing machine using the machine description language from [1]. The volume of our Turing machine is 7x12x26, meaning that the Builder machine from [1] cannot build it. However, because the Builder from [1] can build a larger version of itself, this is not a problem. Hence, together with the mechanism from reference [1] and a larger version of the Builder, we have a way for a Turing machine to self-replicate. What remains to be shown is universality.

.

## Generalization

To demonstrate universality, we must adjust our simplistic notation to a more formal form. In this form, the Turing machine from Table 4 is shown in Table 5. Here, q1 corresponds to the above R-state and q2 corresponds to the above L-state. The rules are to be read in the following way. For instance, rule #1 reads as follows: if we are in state q1 and read symbol 0, then we write symbol 0, move to the right, and change to the new state q1.

```
Symbols: 0, 1, #
States: q1, q2
Rules:
1.  q1 0 0 R q1
2.  q1 1 1 R q1
3.  q1 # # R q2
4.  q2 0 1 H -
5.  q2 1 0 L q2
6.  q2 # 1 H -

Default symbol: #
Initial state: q1
Tape: ...##01##...
Head position: to the left of the first 0
```

Table 5: Transition table for binary incrementer

.

## Adding two Unary Numbers

To gain more familiarity with the new notation, and also to show a textual analog of the graphical notation, let us consider another example: a Turing machine for adding two unary numbers, as shown in Fig. 18.

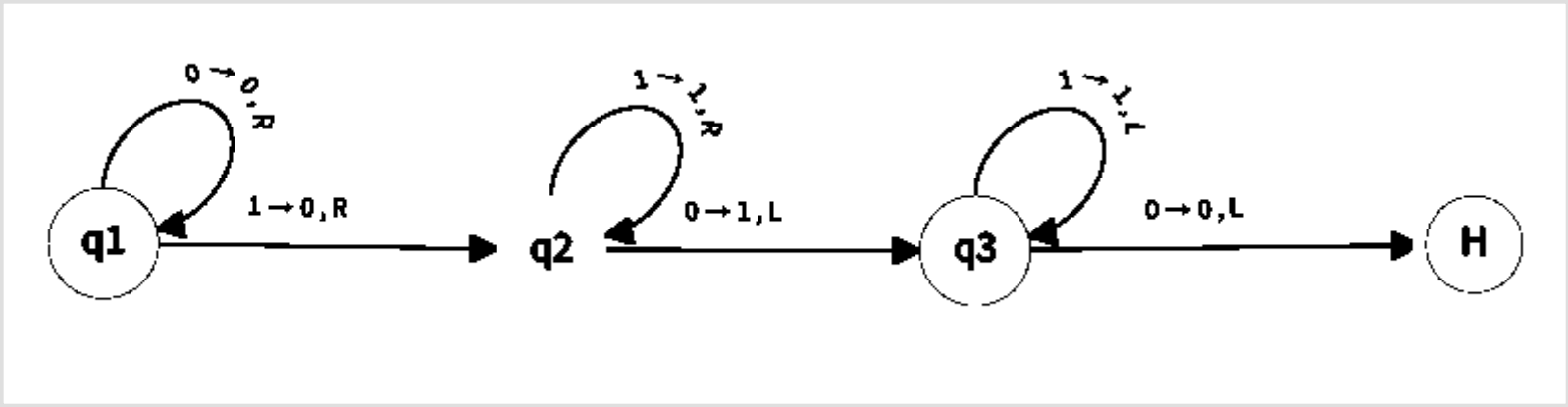

a

Figure 18: Unary adder

Table 6 presents a formal description of this Turing machine. It has the symbols 0 and 1, and the three states q1, q2, and q3. The tape contains two unary numbers separated by a 0. The rest of the tape is all zeros. To add these two numbers, all we need to do, turn the first 1 into a 0, and the 0 in the middle into a 1. The Turing machine in Table 6 with the given rules does exactly this.

```
Symbols: 0, 1
States: q1, q2, q3
Rules:
1.  q1 0 0 R q1
2.  q1 1 0 R q2
3.  q2 1 1 R q2
4.  q2 0 1 L q3
5.  q3 1 1 L q3
6.  q3 0 0 H -

Default symbol: 0
Initial state: q1
Tape: ...0001101000...
Head position: to the left of the first 1
```

Table 6: Transition table for unary adder

This is a Turing machine with three states and two symbols, that is TM(3,2). Since for states odd codons are allowed, we can use 2-codons for the symbols and 3-codons for the states. We use this example and translate it into our mechanism. However, instead of using a graphical representation, as shown in Figs. 2 and 3, we use a textual representation. Then, the string '01' corresponds to the 2-codon from Fig. 3(a) and the string '10' corresponds to the 2-codon from Fig. 3(b). Simply put, the presence of a block is represented by a 1 and the absence by a 0, and support structures are ignored. With this convention, we follow the following steps:

**Step 1 (Defining symbols and states):** In Step 1 of the conversion process, we define the symbols in Table 7, where the read representation is the opposite of the write representation,

```
    write  read
0:   01     10
1:   10     01
```

Table 7: List of symbols

and the states in Table 8, where the read representation is again the opposite of the write representation.

```
     write  read
q1:  001    110
q2:  010    101
q3:  100    011
qH:  111    000
```

Table 8: List of states

.

**Step 2 (Preparing the initial tape):** This representation is then used to translate the given tape '011010' into our mechanism. We replace the symbols with their corresponding write versions, that is, symbol 0 by '01', and symbol 1 by '10':

**..._01_10_10_01_10_01_...**

Then between each symbol we place the write version of the halt state:

**..._111_01_111_10_111_10_111_01_111_10_111_01_111_...**

Finally, we initiate the left-most state with the initial state q1:

**..._001_01_111_10_111_10_111_01_111_10_111_01_111_...**

This is an equivalent representation to Fig. 11.

.

**Step 3 (Translation rules):** The final step is to translate the rules into the equivalent of tRNA, as shown in Figure 12. As an example, let us consider rule #1. We are in state q1, and assume that it is a right-moving state. This means that we can assume the state to be to the left, which is represented by '110_xx_000'. Furthermore, it reads symbol 0 from the tape, that is '10' according to Table 7, and thus the read part of rule #1 is represented by '110_10_000'. Because it is a right-moving state, the middle part of the tRNA (Fig. 12) will have no hole; hence, it is represented by '111_11_111'. Finally, we arrive at the write part. The rule says that we stay in q1; looking at the write part of Table 8, this is '001'. In addition, rule #1 states that we do a right-move; hence, this is represented by '111_xx_001'. The output symbol to be written is 0, and looking at the write column of Table 7, this is '01', and hence, the write row is '111_01_001'. With this it is straight-forward to translate the rules given in the rules table in Table 6.

```
1. q1 0 0 R q1:
read:  110_10_000
R/L:   111_11_111
write: 111_01_001

2. q1 1 0 R q2:
read:  110_01_000
R/L:   111_11_111
write: 111_01_010

3. q2 1 1 R q2:
read:  101_01_000
R/L:   111_11_111
write: 111_10_010

4. q2 0 1 L q3:
read:  101_10_000
R/L:   011_11_111
write: 100_10_111

5. q3 1 1 L q3:
read:  000_01_011
R/L:   011_11_111
write: 100_10_111

6. q3 0 0 H -:
read:  000_10_011
R/L:   111_11_111
write: 111_01_111
```

Table 9: tRNA / transition rules

.

Let us apply these rules to our tape. This is a textual version of the graphical process described in Fig. 13. At the current position of the head, we search for a matching rule from Table 9 using the read part. Once that is found, we replace that part of the tape with the write part. Finally, depending on whether the R/L has a hole, we move to the left or right.

```
...001_01_111_10_111_10_111_01_111_10_111_01_111_...
   110_10_000
   111_11_111
   111_01_001

...111_01_001_10_111_10_111_01_111_10_111_01_111_...
          110_01_000
          111_11_111
          111_01_010

...111_01_111_01_010_10_111_01_111_10_111_01_111_...
                 101_01_000
                 111_11_111
                 111_10_010

...111_01_111_01_111_10_010_01_111_10_111_01_111_...
                        101_10_000
                        011_11_111
                        100_10_111

...111_01_111_01_111_10_100_10_111_10_111_01_111_...
                 000_01_011
```

```
                 011_11_111
                 100_10_111

...111_01_111_01_100_10_111_10_111_10_111_01_111_...
          000_10_011
          111_11_111
          111_01_111

...111_01_111_01_111_10_111_10_111_10_111_01_111_...
```

The machine halts, and we obtain the result we expect, '001110', the unary number three. All states are in the halt state '111'.

In Step 3 above, we assumed that initially, we are in a right-moving state. For the unary adder, this is true, but it is not true in general. However, a generic solution to this problem exists. For this purpose, consider rule #1 in Table 9.

```
1. q1 0 0 R q1:
read:  110_10_000
R/L:   111_11_111
write: 111_01_001
```

Clearly, there is no ambiguity regarding the output part: the head will move to the right, and thus, the R/L and write parts will be fixed. However entering, we do not know whether the previous state was right-moving or left-moving. If it was right-moving, then the state information is to be found to the left; however, if it was left-moving, it is to be found on the right. Hence, if we were to turn the one rule

```
1. q1 0 0 R q1:
```

into two tRNA, that is

```
read:  110_10_000
R/L:   111_11_111
write: 111_01_001
```

and

```
read:  000_10_110
R/L:   111_11_111
write: 111_01_001
```

then this would solve our ambiguity, and it would not matter if the previous state was left- or right-moving. Having this ambiguity, and needing to avoid it, comes at the cost of needing twice as many tRNA. For the unary adder this is not needed, but for the universal Turing machines that we are about to encounter next, it is.

.

## Universality

To demonstrate universality of our Turing machine, we must show that we can implement one of the known universal Turing machines [6-8]. Of interest to us is the length of the tRNA, which is two times the length of the state plus the length of the symbol, Table 10 lists several known universal Turing machines under this aspect. Recall that the codons for symbols must be even.

| | states | symbols | length |
|---|---|---|---|
| WUTM(2,3) | 2 | 4 | 8 |
| UTM(2,18) | 2 | 6 | 10 |
| UTM(3,10) | 3 | 6 | 12 |
| UTM(4,6) | 4 | 4 | 12 |
| UTM(5,5) | 4 | 4 | 12 |
| UTM(6,4) | 4 | 4 | 12 |

Table 10: List of UTMs

From a length point of view, the WUTM(2,3) of Smith [6] or the UTM(2,18) of Rogozhin [7] may be the most interesting. However, we chose to implement the UTM(5,5) proposed by Neary and Woods [8], as shown in Fig. 19.

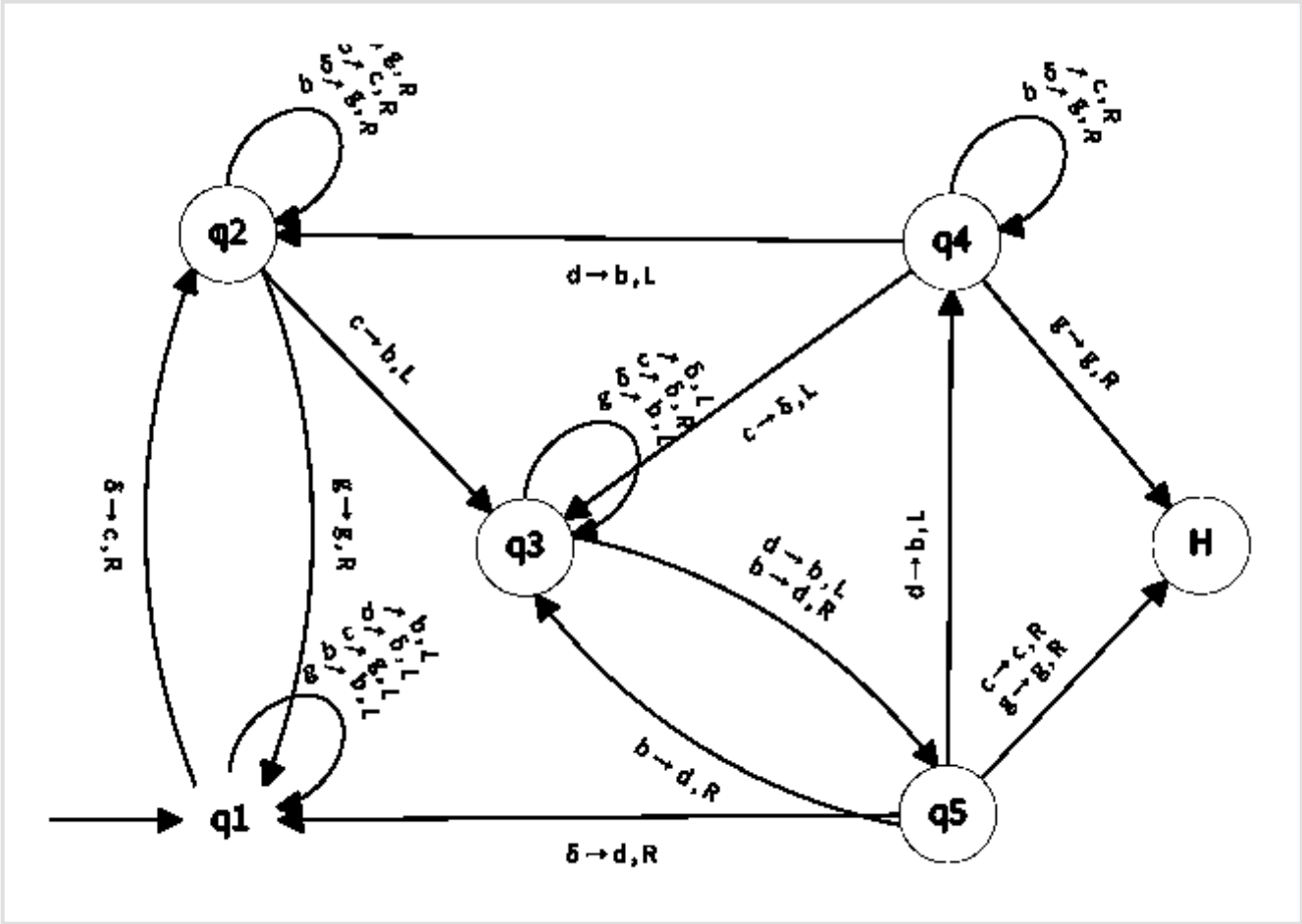

a

Figure 19: UTM(5,5) of Neary and Woods [8]

.

It has five states: q1 through q5, and five symbols: g, b, δ, c, and d, where c is the blank symbol and q1 is the initial state. This information is summarized in Table 11.

**Symbols: g, b, δ, c, d**
**States: q1, q2, q3, q4, q5**
**Transition table:**

| | g | b | δ | c | d |
|---|---|---|---|---|---|
| q1 | b L q1 | g L q1 | c R q2 | δ L q1 | b L q1 |
| q2 | g R q1 | g R q2 | c R q2 | b L q3 | g R q2 |
| q3 | b L q3 | d R q5 | δ R q3 | δ L q3 | b L q5 |
| q4 | | g R q4 | c R q4 | δ L q3 | b L q2 |
| q5 | | d R q3 | d R q1 | | b L q4 |

**Default symbol: c**
**Initial state: q1, right moving**
**Tape: ...dddddδδδbbbδbbbbbbbδcccccc...**
**Head position: to the left of the last of the three δ**

Table 11: Transition table for UTM(5,5) from Neary and Woods [8]

We need to convert the above transition table into our longer rules form, Table 12, with small additions for the undefined rules:

```
 1.  q1 g b L q1
 2.  q1 b g L q1
 3.  q1 δ c R q2
 4.  q1 c δ L q1
 5.  q1 d b L q1
 6.  q2 g g R q1
 7.  q2 b g R q2
 8.  q2 δ c R q2
 9.  q2 c b L q3
10.  q2 d g R q2
11.  q3 g b L q3
12.  q3 b d R q5
13.  q3 δ δ R q3
14.  q3 c δ L q3
15.  q3 d b L q5
16.  q4 g g H -
17.  q4 b g R q4
18.  q4 δ c R q4
19.  q4 c δ L q3
20.  q4 d b L q2
21.  q5 g g H -
22.  q5 b d R q3
23.  q5 δ d R q1
24.  q5 c c H -
25.  q5 d b L q4
```

Table 12: Transition rules for UTM(5,5) from Neary and Woods [8]

We follow the steps outlined above.

**Step 1 (Defining symbols and states):** In Step 1 of the conversion process, we define the symbols (Table 13),

```
   write
g: 000111
b: 001011
δ: 010011
c: 100011
d: 001110
```

Table 13: List of symbols for UTM(5,5)

and the states (Table 14), where we only list the write part and omit the read part, because it is simply the opposite.

```
    write
q1: 000111
q2: 001011
q3: 010011
q4: 100011
q5: 001110
qH: 111111
```

Table 14: List of states for UTM(5,5)

.

**Step 2 (Preparing the initial tape):** We use this representation to translate the given tape 'δδ|**b**|bb' into our mechanism. We replace the symbols with their corresponding write versions according to Table 13, that is, the symbol δ by '010011', and the symbol b by '001011', etc.:

```
..._010011_010011_|_001011_|_001011_001011...
```

Then, between each symbol, we place the write version ('111111') of the halt state:

```
..._111111_010011_111111_010011_|_001011_111111_001011_111111_001011_111111_...
```

and finally, we initiate the left state next to our head position with our initial state q1, that is '000111':

```
..._111111_010011_111111_010011_000111_001011_111111_001011_111111_001011_111111_...
```

This is an equivalent representation to Fig. 11.

.

**Step 3 (Translation rules):** The final step is to translate the rules into the equivalent of tRNA, as shown in Figure 12. As indicated, in the case of UTM(5,5) we have to worry about the ambiguity of not knowing for certain if the previous state was left or right moving, and in general, we have to assume both. The solution is to have two tRNA for every transition rule from Table 12, resulting in a total of 50 tRNA. The first four are listed in Table 15 and the rest can be found in Appendix B.

```
1. q1 g b L q1:
read:  '111000'_111000_'000000'
read1: 111000_111000_000000
read2: 000000_111000_111000
R/L:   011111_111111_111111
write: 000111_001011_111111

2. q1 b g L q1:
read:  '111000'_110100_'000000'
read1: 111000_110100_000000
read2: 000000_110100_111000
R/L:   011111_111111_111111
write: 000111_000111_111111

...
```

Table 15: tRNA / transition rules for UTM(5,5)

.

Applying these rules to the given tape results in the following calculation:

```
dddddδδδbbbδbbbbbbbδcccccc
dddddδδδgbbδbbbbbbbδcccccc
dddddδδcgbbδbbbbbbbδcccccc

... 86 steps ...

dggbbδδδbbbδbbbbbbbδbbcccc
dgbbbδδδbbbδbbbbbbbδbbcccc
dbbbbδδδbbbδbbbbbbbδbbcccc
```

Appendix A provides the omitted lines. This result is the same as that obtained by Neary and Woods [8] and corresponds to their Cycle 1 presented on p.118. Hence, we have shown that our mechanism can emulate the UTM(5,5) of Neary and Woods and is thus universal.

We have shown that our Turing machine can self-replicate with the mechanism from [1], but we are confident that the mechanism from [2] can also be used.

.

## Discussion

The key contributions of this work include the realization that the Decoder machine from [1] can function as a finite-state machine, formalization of a Turing Machine capable of self-replication, and demonstration of universality via the UTM(5,5) model. This marks a step forward in the field of molecular computing and mechanical self-replication, and provides new insights into how such systems can be designed.

The use of the Decoder machine as a finite-state machine capable of performing basic computational tasks may come as a surprise, but it lies in the fact that the tRNA are basically lookup tables, wherein the true computational power lies, and clearly shows that computational power is at the heart of self-replicating machines, that is biological cells.

The finite-state machine (FSM) essentially operates in the same way as the Decoder, which mirrors the function of the ribosome in biological systems, with the primary difference being the type of tRNA involved. Thus, if a FSM were advantageous to a living cell, evolution could have easily produced one. This raises an interesting question: can a FSM serve a useful function within a cell?

Finite-state machines are very useful in various computational tasks such as counting, basic forms of computation, parsing, pattern recognition, controlling, verification, and modeling. For instance, a simple FSM could allow a cell to generate periodic signals or track how frequently a certain event occurred. In the context of cellular communication, cells or organelles within a cell could potentially exchange short RNA sequences as a means of communicating information, which FSMs could then read and interpret. Such FSMs could also be applied to biological processes such as building complex structures. For example, in organisms that construct shells, FSMs might be used to create periodic or patterned designs. The concept of mechanical computation could prove valuable at the cellular level, similar to how cells respond to sensor inputs or store and process information. This flexibility represents the essence of computing, that is, the ability to generate entirely new outcomes based on external conditions. Unlike simple biological replication or mutation, computational systems allow responses to sensory inputs and dynamic problem-solving.

However, when considering more complex machines such as Turing machines (TM), we need to assess their practicality in biological contexts. The TM proposed in this work is one orders of magnitude more efficient than that described in [3]. However, an important question remains: for an organism, is a universal Turing machine (UTM) truly necessary, or would a specialized FSM suffice? The UTM presented here is relatively inefficient, requiring approximately 50 tRNA, meaning that each computational step involves 50 comparisons on average. A simpler TM design with only five to ten tRNA, focused on performing a specific task, could offer greater efficiency and better serve biological purposes.

We believe that a UTM is not particularly useful within living cells. UTMs are inherently slow, and mechanical computation cannot match the speed of electronic computations. Mimicking a von Neumann architecture, for instance, would be far more time efficient than relying on a UTM for computation.

.

## Conclusion

This study presents a novel approach for constructing self-replicating Turing machines and finite-state machines using a bio-inspired framework grounded in RNA codons and tRNA-matching mechanisms. While the physical implementation of TMs is not new, this work advances the field by reducing the design to two basic building blocks, Mover and Simple blocks, with the former having specific timing requirements. This simplification renders these mechanisms more practical and scalable, particularly in the context of self-replicability. Our findings show that computation is well within reach of biological cells, demonstrating how easily FSMs can be implemented with such simple components.

In addition, we highlight the potential of this bio-inspired system to perform computations in environments that lack conventional manufacturing resources, underscoring its applicability in scenarios where traditional silicon-based technologies are impractical. This study opens new avenues for integrating synthetic biology with programmable computational logic, offering promising applications in areas such as space exploration, where autonomous mechanical computation is crucial. FSMs could be used to regulate simple mechanisms for computation and control in environments such as asteroid mining, where silicon foundries are unavailable. For example, Drexler [13], Freitas and Merkle [14], and Langford et al. [15] suggested that such results are relevant for operations in remote resource-constrained environments that lack existing technological infrastructure. The simplicity and scalability of these bio-inspired systems make them ideal candidates for space exploration and planetary colonization, where computational systems may need to be built from alternative materials and substrates other than silicon.

Several directions have emerged for future research. One potential avenue involves the development of cellular automata machines [16,17] by combining Decoder and Copier machines. This allows us to explore the relationship between computability and buildability. Another area of interest are the Lindenmayer systems [18], which merge computing and building, particularly in the context of folding, offering further possibilities for applying these mechanisms in synthetic biology. Additionally, 2-tag machines [19-21] may present an alternative to universal Turing machines by providing a simpler, more robust solution for universality without the complexity of arbitrary back-and-forth movement of the head of the TM. These machines, which progress in one direction and then reset, could offer a more practical approach to achieving reprogrammability. Finally, there is the potential for exploring von Neumann-type CPU designs [22] for mechanical computation, providing insights into how mechanical CPUs can be realized for real-world applications.

For me, the most intriguing result of this research is the Decoder, which shows how building and computing, how life, self-replication, and computing are related: it is the same machine, the Decoder, that does computing in the FSM/TM and building proteins in the folding mechanism. Essential are the tRNA, the lookup-tables, that represent the program for the FSM/TM, but for building they are used for translating the RNA code into block type.

.

# Appendix

**A)** Result of running UTM(5,5) on given input tape:

```
dddddδδδbbbδbbbbbbbδcccccc
dddddδδδgbbδbbbbbbbδcccccc
dddddδδcgbbδbbbbbbbδcccccc
dddddδδcgbbδbbbbbbbδcccccc
dddddδδcggbδbbbbbbbδcccccc
dddddδδcbgbδbbbbbbbδcccccc
dddddδδδbgbδbbbbbbbδcccccc
dddddδcδbgbδbbbbbbbδcccccc
dddddδccbgbδbbbbbbbδcccccc
dddddδccggbδbbbbbbbδcccccc
dddddδccggbδbbbbbbbδcccccc
dddddδccgggδbbbbbbbδcccccc
dddddδccgbgδbbbbbbbδcccccc
dddddδccbbgδbbbbbbbδcccccc
dddddδcδbbgδbbbbbbbδcccccc
dddddδδδbbgδbbbbbbbδcccccc
dddddcδδbbgδbbbbbbbδcccccc
dddddccδbbgδbbbbbbbδcccccc
dddddcccbbgδbbbbbbbδcccccc
dddddcccgbgδbbbbbbbδcccccc
dddddcccgggδbbbbbbbδcccccc
dddddcccgggδbbbbbbbδcccccc
dddddcccgggcbbbbbbbδcccccc
dddddcccgggcgbbbbbbδcccccc
dddddcccgggcggbbbbbδcccccc
dddddcccgggcgggbbbbδcccccc
dddddcccgggcggggbbbδcccccc
dddddcccgggcgggggbbδcccccc
dddddcccgggcggggggbδcccccc
dddddcccgggcgggggggδcccccc
dddddcccgggcgggggggccccccc
dddddcccgggcgggggggcbccccc
dddddcccgggcgggggggδbccccc
dddddcccgggcggggggbδbccccc
dddddcccgggcgggggbbδbccccc
dddddcccgggcggggbbbδbccccc
dddddcccgggcgggbbbbδbccccc
dddddcccgggcggbbbbbδbccccc
dddddcccgggcgbbbbbbδbccccc
dddddcccgggcbbbbbbbδbccccc
dddddcccgggδbbbbbbbδbccccc
dddddcccggbδbbbbbbbδbccccc
dddddcccgbbδbbbbbbbδbccccc
dddddcccbbbδbbbbbbbδbccccc
dddddccδbbbδbbbbbbbδbccccc
dddddcδδbbbδbbbbbbbδbccccc
. . .
```

```
. . .
dddddδδδbbbδbbbbbbbδbccccc
ddddbδδδbbbδbbbbbbbδbccccc
dddbbδδδbbbδbbbbbbbδbccccc
ddbbbδδδbbbδbbbbbbbδbccccc
dgbbbδδδbbbδbbbbbbbδbccccc
dggbbδδδbbbδbbbbbbbδbccccc
dgggbδδδbbbδbbbbbbbδbccccc
dggggδδδbbbδbbbbbbbδbccccc
dggggcδδbbbδbbbbbbbδbccccc
dggggccδbbbδbbbbbbbδbccccc
dggggcccbbbδbbbbbbbδbccccc
dggggcccgbbδbbbbbbbδbccccc
dggggcccggbδbbbbbbbδbccccc
dggggcccgggδbbbbbbbδbccccc
dggggcccgggcbbbbbbbδbccccc
dggggcccgggcgbbbbbbδbccccc
dggggcccgggcggbbbbbδbccccc
dggggcccgggcgggbbbbδbccccc
dggggcccgggcggggbbbδbccccc
dggggcccgggcgggggbbδbccccc
dggggcccgggcggggggbδbccccc
dggggcccgggcgggggggδbccccc
dggggcccgggcgggggggcbccccc
dggggcccgggcgggggggcgccccc
dggggcccgggcgggggggcgbcccc
dggggcccgggcgggggggcbbcccc
dggggcccgggcgggggggδbbcccc
dggggcccgggcggggggbδbbcccc
dggggcccgggcgggggbbδbbcccc
dggggcccgggcggggbbbδbbcccc
dggggcccgggcgggbbbbδbbcccc
dggggcccgggcggbbbbbδbbcccc
dggggcccgggcgbbbbbbδbbcccc
dggggcccgggcbbbbbbbδbbcccc
dggggcccgggδbbbbbbbδbbcccc
dggggcccggbδbbbbbbbδbbcccc
dggggcccgbbδbbbbbbbδbbcccc
dggggcccbbbδbbbbbbbδbbcccc
dggggccδbbbδbbbbbbbδbbcccc
dggggcδδbbbδbbbbbbbδbbcccc
dggggδδδbbbδbbbbbbbδbbcccc
dgggbδδδbbbδbbbbbbbδbbcccc
dggbbδδδbbbδbbbbbbbδbbcccc
dgbbbδδδbbbδbbbbbbbδbbcccc
dbbbbδδδbbbδbbbbbbbδbbcccc
```

.

.

**B)** tRNA / transition rules for UTM(5,5):

```
1. q1 g b L q1:
read:  '111000'_111000_'000000'
read1: 111000_111000_000000
read2: 000000_111000_111000
R/L:   011111_111111_111111
write: 000111_001011_111111

2. q1 b g L q1:
read:  '111000'_110100_'000000'
read1: 111000_110100_000000
read2: 000000_110100_111000
R/L:   011111_111111_111111
write: 000111_000111_111111

3. q1 δ c R q2:
read:  '111000'_101100_'000000'
read1: 111000_101100_000000
read2: 000000_101100_111000
R/L:   111111_111111_111111
write: 111111_100011_001011

4. q1 c δ L q1:
read:  '111000'_011100_'000000'
read1: 111000_011100_000000
read2: 000000_011100_111000
R/L:   011111_111111_111111
write: 000111_010011_111111

5. q1 d b L q1:
read:  '111000'_110001_'000000'
read1: 111000_110001_000000
read2: 000000_110001_111000
R/L:   011111_111111_111111
write: 000111_001011_111111

6. q2 g g R q1:
read:  '110100'_111000_'000000'
read1: 110100_111000_000000
read2: 000000_111000_110100
R/L:   111111_111111_111111
write: 111111_000111_000111

7. q2 b g R q2:
read:  '110100'_110100_'000000'
read1: 110100_110100_000000
read2: 000000_110100_110100
R/L:   111111_111111_111111
write: 111111_000111_001011

8. q2 δ c R q2:
read:  '110100'_101100_'000000'
read1: 110100_101100_000000
read2: 000000_101100_110100
R/L:   111111_111111_111111
write: 111111_100011_001011

9. q2 c b L q3:
read:  '110100'_011100_'000000'
read1: 110100_011100_000000
read2: 000000_011100_110100
R/L:   011111_111111_111111
write: 010011_001011_111111

10. q2 d g R q2:
read:  '110100'_110001_'000000'
read1: 110100_110001_000000
read2: 000000_110001_110100
R/L:   111111_111111_111111
write: 111111_000111_001011

11. q3 g b L q3:
read:  '101100'_111000_'000000'
read1: 101100_111000_000000
read2: 000000_111000_101100
R/L:   011111_111111_111111
write: 010011_001011_111111

12. q3 b d R q5:
read:  '101100'_110100_'000000'
read1: 101100_110100_000000
read2: 000000_110100_101100
R/L:   111111_111111_111111
write: 111111_001110_001110

13. q3 δ δ R q3:
read:  '101100'_101100_'000000'
read1: 101100_101100_000000
read2: 000000_101100_101100
R/L:   111111_111111_111111
write: 111111_010011_010011

14. q3 c δ L q3:
read:  '101100'_011100_'000000'
read1: 101100_011100_000000
read2: 000000_011100_101100
R/L:   011111_111111_111111
write: 010011_010011_111111

15. q3 d b L q5:
read:  '101100'_110001_'000000'
read1: 101100_110001_000000
read2: 000000_110001_101100
R/L:   011111_111111_111111
write: 001110_001011_111111

16. q4 g g H -:
read:  '011100'_111000_'000000'
read1: 011100_111000_000000
read2: 000000_111000_011100
R/L:   111111_111111_111111
write: 111111_000111_111111
```

.

```
17. q4 b g R q4:
read:  '011100'_110100_'000000'
read1: 011100_110100_000000
read2: 000000_110100_011100
R/L:   111111_111111_111111
write: 111111_000111_100011

18. q4 δ c R q4:
read:  '011100'_101100_'000000'
read1: 011100_101100_000000
read2: 000000_101100_011100
R/L:   111111_111111_111111
write: 111111_100011_100011

19. q4 c δ L q3:
read:  '011100'_011100_'000000'
read1: 011100_011100_000000
read2: 000000_011100_011100
R/L:   011111_111111_111111
write: 010011_010011_111111

20. q4 d b L q2:
read:  '011100'_110001_'000000'
read1: 011100_110001_000000
read2: 000000_110001_011100
R/L:   011111_111111_111111
write: 001011_001011_111111

21. q5 g g H -:
read:  '110001'_111000_'000000'
read1: 110001_111000_000000
read2: 000000_111000_110001
R/L:   111111_111111_111111
write: 111111_000111_111111

22. q5 b d R q3:
read:  '110001'_110100_'000000'
read1: 110001_110100_000000
read2: 000000_110100_110001
R/L:   111111_111111_111111
write: 111111_001110_010011

23. q5 δ d R q1:
read:  '110001'_101100_'000000'
read1: 110001_101100_000000
read2: 000000_101100_110001
R/L:   111111_111111_111111
write: 111111_001110_000111

24. q5 c c H -:
read:  '110001'_011100_'000000'
read1: 110001_011100_000000
read2: 000000_011100_110001
R/L:   111111_111111_111111
write: 111111_100011_111111

25. q5 d b L q4:
read:  '110001'_110001_'000000'
read1: 110001_110001_000000
read2: 000000_110001_110001
R/L:   011111_111111_111111
write: 100011_001011_111111
```

.

**C)** Detailed description (MDL) of Turing machine Fig. 14(b):

| | | |
|---|---|---|
| // z=0:<br><br>M25 | // z=1:<br><br>M29<br><br>M25<br><br>M29 | // z=2:<br><br>M29<br><br>M25<br><br>M29 |
| // z=3:<br><br>M29<br><br>M25<br><br>M29 | // z=4:<br><br>M29<br><br>M25<br><br>M29 | // z=5:<br><br>M18<br>M18<br>M18<br>M18<br><br>M44<br>M44<br>M44<br>M44<br><br>M48<br>M48<br>M48<br>M48 |

```
// z=6:
```

```
// z=7:
```

```
// z=8:
M15
M15
M15
M15
M15
M15
M15
M15
M15
M15
M15
M15
M06M06
```

```
// z=9:
b__














               M40
      M03      M40
      M42      M40
      M42      M40M31
      M42         b__
      M42   b__b__b__
      M42   b__
      M42   b__
      b__   b__
      b__   b__
      b__b__b__
```

```
// z=10:
b__b__b__
      b__
      b__
      b__
      b__
      b__
      b__
      b__
      b__
      b__
      b__
      b__
      b__
      b__
      b__
      b__
      b__
   b__b__
      b__
      b__
      b__
      b__
```

```
// z=11:

















   b__
```